\documentclass[aps,twocolumn,pre]{revtex4}
\usepackage{amssymb,epsfig}
\newcommand{\cbeta}{{\cal B}}

\begin{document}
\draft   
\title{Classification of phase transitions of finite Bose-Einstein
condensates in power law traps by Fisher zeros}
\author{Oliver M{\"u}lken, Peter Borrmann, Jens Harting, and 
Heinrich Stamerjohanns}
\affiliation{Department of Physics, Carl von Ossietzky University
Oldenburg, D-26111 Oldenburg, Germany}
\date{\today}
\begin{abstract}
We present a detailed description of a classification scheme for phase
transitions in finite systems based on the distribution of Fisher zeros of the
canonical partition function in the complex temperature plane.  We apply this
scheme to finite Bose-systems in power law traps within a semi-analytic
approach with a continuous one-particle density of states $\Omega(E)\sim
E^{d-1}$ for different values of $d$ and to a three dimensional harmonically
confined ideal Bose-gas with discrete energy levels. Our results indicate that
the order of the Bose-Einstein condensation phase transition sensitively
depends on the confining 
potential. 
\end{abstract}
\pacs{PACS numbers: 05.20.-y, 05.30.Jp, 64.60.-i}
\maketitle
\section{Introduction}
In 1924 S.~Bose and A.~Einstein predicted that in a system of bosons at
temperatures below a certain critical temperature $T_C$  the single-particle
ground state is macroscopically occupied~\cite{Bose1924a}. This effect is
commonly referred as Bose-Einstein condensation and a large number of
phenomena, among others the condensation phenomena in alkali atoms, the
superfluidity of $^4$He and the superconductivity, are identified as signatures
of this effect. However, the physical situation is very intricate in most
experiments. 

Recent experiments with dilute gases of alkali atoms in
magnetic~\cite{Anderson1995a} and optical~\cite{Stamper-Kurn} traps are in some
sense the up to now best experimental approximation of the ideal
non-interacting Bose-Einstein system in an external power law potential. The
achievement of ultra-low temperatures by laser cooling and evaporative cooling
opens the opportunity to study the Bose-Einstein condensation under systematic
variation of adjustable external parameters, e.g the trap geometry, the number
of trapped atoms, the temperature, and by the choice of the alkali atoms the
effective interparticle interactions. Even in the approximation of
non-interacting particles the explanation of these experiments requires some
care, because the number of bosons in these novel traps is finite and fixed and
the standard grand-canonical treatment is not appropriate. The effect of the
finite particle numbers on the second moments of the distribution function,
e.g. the specific heat and the fluctuation of  the ground state occupation
number has been addressed in a number of publications
\cite{recurnew,Grossmann1997a}.  In \cite{recurnew,recurold} we have presented
a recursion method to calculate the canonical partition function for
non-interacting bosons and investigated the dependency of the thermodynamic
properties of the condensate on the trap geometry.

The order of the phase transition in small systems sensitively depends on
finite size effects. Compared to the macroscopic system even for as
simple systems as the 3-dimensional ideal gas the order of the phase
transition might change for mesoscopic systems where the number of
particles is finite or for trapped gases with different trap geometries. 

In this paper we address the classification of the phase transition of a finite
number of non-interacting bosons in a power law trap with an effective
one-particle density of states $\Omega(E)= E^{d-1}$ being formally equivalent
to a $d$-dimensional harmonic oscillator or a $2d$-dimensional ideal gas.  We
use a classification scheme based on the distribution of zeros of the canonical
partition function initially developed by Grossman {\sl
et~al.}~\cite{Gross1967}, and Fisher {\sl et~al.}~\cite{Fisher}, which has been
extended by us~\cite{borrmann99} as a classification scheme for finite systems.
On the basis of this classification scheme we are able to extract a
qualitative difference between the order of the phase transition occuring
in Bose-Einstein condensates in 3-dimensional
traps~\cite{Bagnato1991,Rojas} and in 2-dimensional traps which was
recently discovered by Safonov {\sl et al.} in a gas of hydrogen atoms
absorbed on the surface of liquid helium~\cite{safonov98}. Since we do not
consider particle interactions this difference is only due to the
difference in the confining potential.

We give a detailed review of the classification scheme in Sec.~\ref{Class_sec}.
In Sec.~\ref{results} we present the method for the calculation of the
canonical partition function in the complex plane and describe details of the
numerical implementation. Our results for $d=1-6$ and particle numbers varying
from 10 to 300 are presented in Sec.~\ref{results} as well as calculations for
a 3-dimensional parabolically confined Bose-gas. 

\section{Classification scheme}\label{Class_sec}
In 1952 Yang and Lee have shown that the grand canonical partition function can
be written as a function of its zeros in the complex fugacity plane, which lie
for systems with hard-core interactions and for the Ising model on a unit
circle~\cite{Yang1952a}.

Grossmann {\sl et al.}~\cite{Gross1967} and Fisher~\cite{Fisher} have extended
this approach to the canonical ensemble by analytic continuation of the inverse
temperature to the complex plane $\beta\to\cbeta=\beta+i\tau$.  Within this
treatment all phenomenologically known types of phase transitions in
macroscopic systems can be identified from the properties of the distribution
of zeros of the canonical partition function. 

In \cite{borrmann99} we have presented a classification scheme for finite
systems which has its macroscopic equivalent in the scheme given by Grossmann.
As usual the canonical partition function reads
\begin{equation}
Z(\cbeta) = \int {\rm d}E \ \Omega(E) \ \exp(-\cbeta E),
\end{equation}
which we write as a product $Z(\cbeta)=Z_{\rm lim}(\cbeta) Z_{\rm
int}(\cbeta)$, where $Z_{\rm lim}(\cbeta)$ describes the limiting behavior of
$Z(\cbeta)$ for $T\to\infty$ imposing that $\lim_{T\to\infty} Z_{\rm
int}(\cbeta)=1$. This limiting partition function will only depend on the
external potential applied to the system, whereas $Z_{\rm int}(\cbeta)$ will
depend on the specific interaction between the system particles. E.g.~for a
$N$-particle system in a $d$-dimensional harmonic trap $Z_{\rm
lim}(\cbeta)=\cbeta^{-dN}$ and thus the zeros of $Z(\cbeta)$ are the same as
the zeros of $Z_{\rm int} (\cbeta)$. Since the partition function is an
integral function, the zeros $\cbeta_k = \cbeta_{-k}^* = \beta_k + i \tau_k \
(k \in \mathbb{N})$ are complex conjugated and the partition function reads
\begin{eqnarray}
Z(\cbeta) &=& Z_{\rm lim}(\cbeta) \ Z_{\rm int}(0) \ \exp(\cbeta 
\partial_\cbeta \ln Z_{\rm int}(0)) \nonumber\\ 
&\times& \prod_{k\in\mathbb{N}} \left( 
1-\frac{\cbeta}{\cbeta_k} \right) 
\left( 1-\frac{\cbeta}{\cbeta_k^*} \right) 
\ \exp\left( \frac{\cbeta}{\cbeta_k} + \frac{\cbeta}{\cbeta_k^*} \right).
\end{eqnarray}

The zeros of $Z(\cbeta)$ are the poles of the Helmholtz free energy $F(\cbeta)
= -\frac{1}{\cbeta} \ln Z(\cbeta)$, i.e. the free energy is analytic everywhere
in the complex temperature plane except at the zeros of $Z(\cbeta)$. 

Different phases are represented by regions of holomorphy which are separated
by zeros lying dense on lines in the complex temperature plane. In finite
systems the zeros do not squeeze on lines which leads to a more blurred
separation of different phases.  We interpret the zeros as boundary posts
between two phases.  The distribution of zeros contains the complete
thermodynamic information about the system and all thermodynamic properties are
derivable from it. Within this picture the interaction part of the specific
heat is given by
\begin{equation}
C_{V,{\rm int}}(\cbeta)=-k_{\rm B}\cbeta^2 \ \sum_{k\in\mathbb{N}}
\left[\frac{1}{(\cbeta_k-\cbeta)^2}+\frac{1}{(\cbeta_k^*-\cbeta)^2}\right].
\label{specheat}
\end{equation} 
The zeros of the partition function are  poles of $C_V(\cbeta)$. As can be seen
from Eq.~(\ref{specheat}) a zero approaching the real axis infinitely close
causes a divergence at real temperature. The contribution of a zero $\cbeta_k$
to the specific heat decreases with increasing imaginary part $\tau_k$. Thus,
the thermodynamic properties of a system are governed by the zeros of $Z$ close
to the real axis.

The basic idea of the classification scheme for phase transitions in small
systems presented in \cite{borrmann99} is that the distribution of zeros close
to the real axis can  approximately be described by three parameters, where two
of them reflect the order of the phase transition and the third merely the size
of the system.

\begin{figure}
\centerline{\epsfig{file=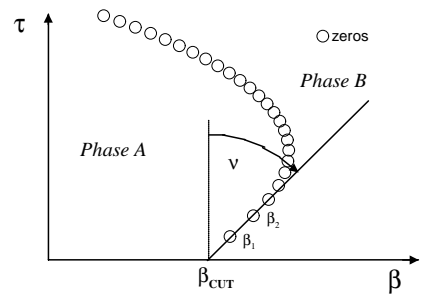,clip=,width=4cm,angle=270}}
\caption{Schematic illustration of the zeros in the complex temperature plane.}
\label{schema}
\end{figure}

We assume that the zeros lie on straight lines (see Fig.~\ref{schema}) with a
discrete density of zeros given by 
\begin{equation}
\phi(\tau_k)=\frac12\left(\frac{1}{|\cbeta_k-\cbeta_{k-1}|}+
\frac{1}{|\cbeta_{k+1}-\cbeta_k|}\right),\label{discretedensity}
\end{equation} 
with $k=2,3,4,\cdots$, and approximate for small $\tau$ the density of zeros by
a simple power law $\phi(\tau)\sim\tau^\alpha$.  Considering only the first
three zeros the exponent $\alpha$ can be estimated as  
\begin{equation}
\alpha=\frac{\ln\phi(\tau_3)-\ln\phi(\tau_2)}{\ln\tau_3-\ln\tau_2}.
\end{equation} 
The second parameter to describe the distribution of zeros is given by
$\gamma=\tan\nu \sim (\beta_2-\beta_1)/ (\tau_2-\tau_1)$ where $\nu$ is the
crossing angle of the line of zeros with the real axis (see Fig.~\ref{schema}).
The {\sl discreteness} of the system is reflected in the imaginary part
$\tau_1$ of the zero closest to the real axis.

In the thermodynamic limit we have always $\tau_1\to0$. In this case the
parameters $\alpha$ and $\gamma$ coincide with those defined by Grossmann {\sl
et al}~\cite{Gross1967}, who have shown how different types of phase
transitions can be attributed to certain values of $\alpha$ and $\gamma$.  They
claimed that $\alpha=0$ and $\gamma=0$ corresponds to a first order phase
transition, second order transitions correspond  to $0<\alpha<1$ with
$\gamma=0$ or $\gamma\neq0$, third order transitions to $1\leq\alpha<2$ with
arbitrary values of $\gamma$, and that all higher order phase transition correspond
to $\alpha>1$.  For macroscopic systems (with $\tau_1\to0$) $\alpha$ cannot be
smaller than zero, because this would cause a divergence of the internal
energy.  However in small systems with a finite $\tau_1$ this is possible.

In our classification scheme we therefore define phase transitions in small
systems to be of first order for $\alpha \leq 0$, while second and higher order
transitions are defined in complete analogy to the Grossmann scheme augmented
by the third parameter $\tau_1$. The definition of a critical temperature
$\beta_C$ in small systems is crucial and ambiguous since no thermodynamic
properties diverge.  Thus, different definitions are possible.  We define the
critical temperature as $\beta_{\rm cut} = \beta_1 - \gamma \tau_1$, i.e. the
crossing point of the approximated line of zeros with the real temperature
axis.  An alternative definition is the real part of the first complex zero
$\beta_1$.  In the thermodynamic limit both definitions coincide. 

\begin{figure}[htb]
\centerline{\epsfig{file=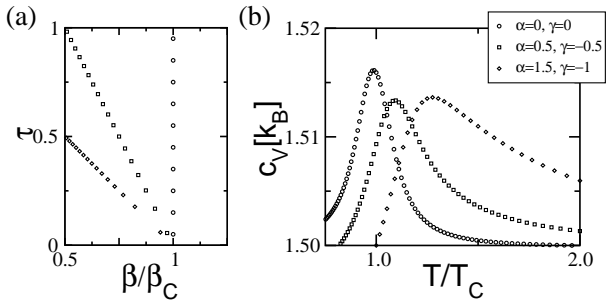,clip=,width=8cm}}
\caption{Plot of (a) generated zeros lying on straight lines to simulate first
($\alpha=0$ and $\gamma=0$), second ($\alpha=0.5$ and $\gamma=-0.5$), and third
($\alpha=1.5$ and $\gamma=-1$) order phase transitions and (b) the appropriate
specific heats per particle.}\label{genzeros}
\end{figure}

Comparing the specific heats calculated for different discrete distributions of
zeros shows the advantages of this classification scheme. Fig.~\ref{genzeros}
shows (a) three distributions of zeros lying on straight lines corresponding to
a first order transition ($\alpha=0$ and $\gamma=0$), a second order transition
($\alpha=0.5$ and $\gamma=-0.5$), and a third order phase transition
($\alpha=1.5$ and $\gamma=-1$) and (b) the pertinent specific heats. In all
cases the specific heat exhibit a hump extending over a finite temperature
region and cannot be used to classify the phase transition.  In contrast, even
for very small systems (large $\tau_1$) the order of the phase transition is
extractable from the distribution of zeros.

The zeros of the canonical partition function have a distinct geometrical
interpretation which explains the smoothed curves of the specific heat and
other thermodynamic properties in finite systems.

\begin{figure}[htb]
\centerline{\epsfig{file=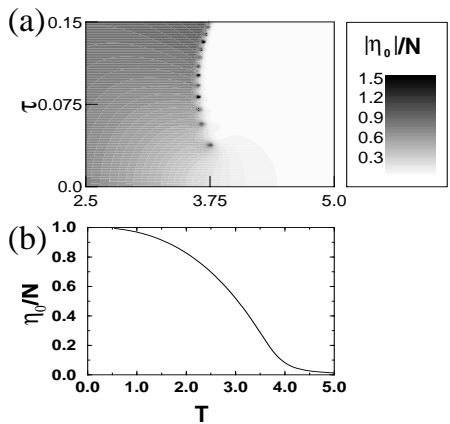,clip=,width=6cm}}
\caption{Comparison of (a) $|\eta_0|/N$ with (b) the appropriate value of $\eta_0$ at
real temperatures for a 120 particle harmonically trapped ideal Bose-gas (note
that $\hbar=k_{\rm B} =\omega =1$).}
\label{discreteoccupn0}
\end{figure}

Fig.~\ref{discreteoccupn0} shows (a) the ground state occupation number
$|\eta_0(\cbeta)|/N$ in the complex temperature plane and (b) the ground
state occupation number at real temperatures  for a finite ideal Bose gas of
$N=120$ particles, where $\eta_0(\cbeta)$ is given
by the derivative of the logarithm of the canonical partition function
$Z(\cbeta)$ with respect to the ground state energy $\epsilon_0$,
i.e.~$\eta_0(\cbeta) = -\frac{1}{\cbeta} \partial_{\epsilon_0}
Z(\cbeta)/Z(\cbeta)$. 

Zeros of the partition function are poles of $\eta_0(\cbeta)$
and are indicated by dark spots, which influence the value of the ground
state occupation number at real temperatures impressively. Every
pole seems to {\sl radiate} onto the real axis and therefore
determines the occupation number at real temperatures. This {\sl
radiation} extends over a broad temperature range so that the
occupation number for real temperatures does not show a
discontinuity but a smoothed curve.  A closer look at
Eq.~(\ref{specheat}) gives the mathematical explanation for this
effect.  The discrete distribution of zeros, i.e. $\tau_1 > 0$,
inhibits the specific heat and all other thermodynamic properties to
show a divergency at some critical temperature because the
denominators of the arguments of the sum remain finite.

Without going into a detailed analysis we note that in the
thermodynamic limit the parameter $\alpha$ is connected to the
critical index for the specific heat by
\begin{equation}
C_V \sim (\beta - \beta_c)^{\alpha-1}.
\end{equation}
However, since critical indices are used to describe the shape of a
divergency at the critical point an extension to small systems seems
to be more or less academical.
 
The introduction of complex temperatures might seem artificial at first
sight but, in fact, the imaginary parts $\tau_k$ of the complex zeros
$\cbeta_k$ have an obvious quantum mechanical interpretation. We write
the quantum mechanical partition function as
\begin{eqnarray}
Z(\beta+i\tau/\hbar)&=&
{\rm Tr}(\exp(-i\tau\hat H/\hbar)\exp(-\beta\hat H)) \label{canstate}\\
&=&\ \left< \Psi_{\rm can} | 
\exp(-i\tau\hat H/\hbar) | \Psi_{\rm can} \right> \\
&=&\  \left< \Psi_{\rm can} (t=0)| \Psi_{\rm can} (t=\tau) \right> ,
\end{eqnarray}
introducing a {\sl canonical state} as a sum over Boltzmann-weighted
eigenstates $\left|\Psi_{\rm can}\right>=\sum_k
\exp(-\beta\epsilon_k/2) \left|\phi_k\right>$. We explicitly write
the imaginary part as $\tau/\hbar$ since the dimension is
$1/\left[{\rm energy}\right]$ and the imaginary part therefore can
be interpreted as time. Then the imaginary parts $\tau_k$ of the
zeros resemble those times for which the overlap of the initial
canonical state with the time evoluted state vanishes. However, they are
not connected to a single system but to a whole ensemble of
identical systems in a heat bath with an initial Boltzmann
distribution.

\section{BEC in power law traps}\label{results}
In this section we assume  a continuous single particle density of states
$\Omega(E)=E^{d-1}$ as an approximation for a  $d$-dimensional harmonic
oscillator or a $2 d$-dimensional ideal gas.  E.g. for the harmonic
oscillator this corresponds to the limit of $\hbar \omega \to 0$ and
taking only the leading term of the degeneracy of the single particle
energy levels.  The one-particle partition function is given by the
Laplace transformation
\begin{equation} \label{contpart}
Z_1(\cbeta)=\int{\rm d} E \ E^{d-1} \exp (-\cbeta E) = (d-1)! \ \cbeta^{-d}.
\end{equation}
The canonical partition function for $N$ non-interacting bosons can be
calculated by the following recursion~\cite{recurold}
\begin{equation}\label{oldrecur}
Z_N(\cbeta) = \frac{1}{N} \ \sum_{k=1}^N \ Z_1(k \cbeta) \ Z_{N-k}(\cbeta),
\end{equation}
where $Z_1(k \cbeta) = \sum_i \exp(-k \cbeta \epsilon_i) $ is the
one-particle partition function at temperature $k \cbeta$ and
$Z_0(\cbeta)=1$.
For small particle numbers this recursion
works fine, even though its numerical effort grows proportional to
$N^2$.  

With (\ref{contpart}) as $Z_1$ Eq.~(\ref{oldrecur}) leads to a
polynomial of order N in $(1/\cbeta)^d$ for $Z_N$ which can be easily
generated using {\sc Maple} or {\sc Mathematica}. The zeros of this
polynomial can be found by standard numerical methods.

\begin{figure}[htb]
\centerline{\epsfig{file=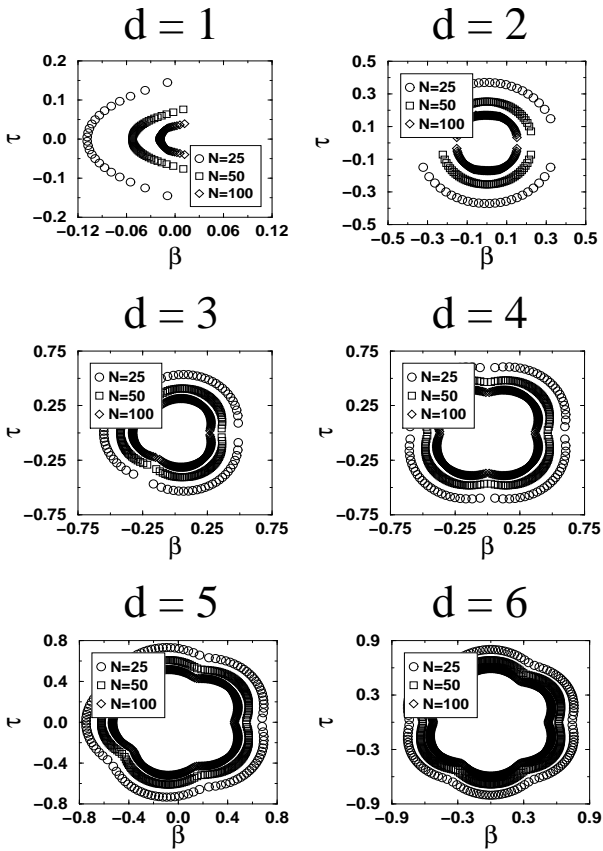,clip=,width=8cm}}
\caption{Distribution of zeros for Bose-Einstein Condensates with continuous
one-particle density of states $\Omega(E)=E^{d-1}$ for $d=1-6$.}\label{analytzeros}
\end{figure}

\begin{figure}[htb]
\centerline{\epsfig{file=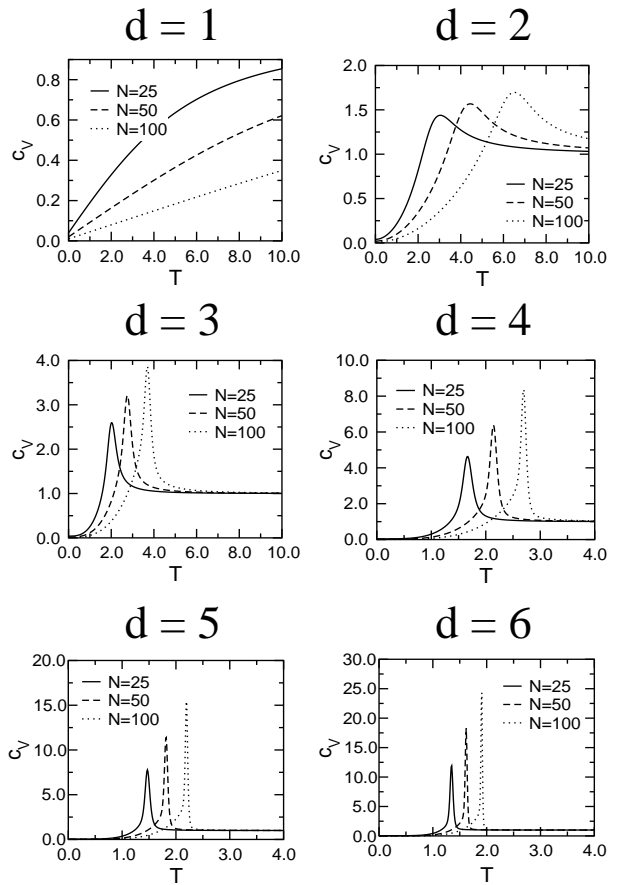,clip=,width=8cm}}
\caption{Specific heat scaled by d$N$ of Bose-Einstein Condensates with
continuous one-particle density of states for $d=1-6$.}\label{analytcv}
\end{figure}

Fig.~\ref{analytzeros} displays the zeros of the $N$-particle
partition function for $d=1-6$ in the complex
temperature plane for particle numbers $N=25, 50$ and $100$. For $d=2-6$ 
the zeros approach the positive real axis with increasing particle number
and are shifted to higher temperatures which is already at first sight an
indicator of phase transitions. For $d=1$ the zeros approach the real
axis only at negative temperature. This behavior is consistent with the 
usual prediction that there is no Bose-Einstein condensation for the
one-dimensional harmonic oscillator
and the two-dimensional ideal Bose gas \cite{Bagnato1991}.

The symmetry of the distributions of zeros is due to the fact that 
$Z_N$ is a polynomial in $\cbeta^{-d}$. For this reason it can be
inferred that for $d \to \infty$ the zeros lie on a perfect circle.

Fig.~\ref{analytcv} shows the corresponding specific heats calculated
using equation (\ref{specheat}). As expected, for $d=1$ the specific
heat has no hump and approaches with increasing temperature the
classical value. We therefore expel the analysis of $d=1$ 
from the discussions below.  For $d=2-6$ the specific heats show
humps or peaks, which get sharper with increasing $d$ and increasing
particle number. However, from these smooth curves the orders of the
phase transition cannot be deduced.

\begin{figure}[htb]
\centerline{\epsfig{file=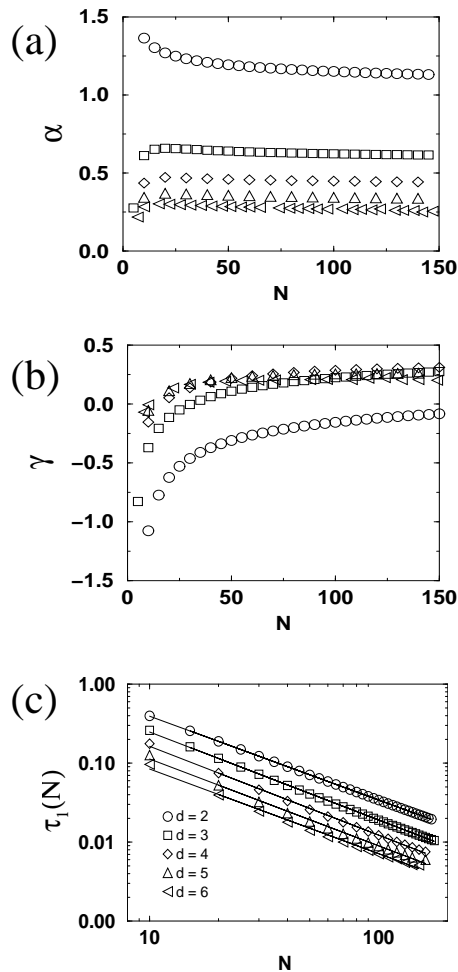,clip=,width=6cm}}
\caption{Classification parameters $\alpha,~\gamma~$ and $\tau_1$ for $d=2-6$
versus particle numbers $N$.}\label{analytdimn}
\end{figure}

\begin{figure}[htb]
\centerline{\epsfig{file=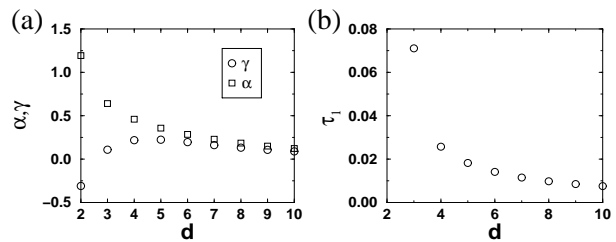,clip=,width=8cm}}
\caption{Classification parameters for $N=50$ for different densities of
states $\Omega(E) = E^{d-1}$ and $d=2-10$.}
\label{analytclassdim}
\end{figure}

In Fig.~\ref{analytdimn} the classification parameters $\alpha,\gamma,\tau_1$
defined above are plotted for two to six dimensions and particle numbers up to
$N=100$.  For all values of $d$ the para\-meter $\alpha$ is only a slightly
varying function of $N$ and approaches very fast an almost constant value.
Since $\alpha$ is the primary classification parameter from
Fig.~\ref{analytdimn}(a) we can directly infer that the $d=2$ system exhibits a
third order phase transition ($\alpha >1)$ while the transition for all higher
dimensions is of second order ($0\leq \alpha \leq 1$). For $N=50$ the
dependence of $\alpha$ on $d$ is plotted in Fig.~\ref{analytclassdim}(a). Since
$\alpha$ decreases rather rapidly with increasing $d$ it can be speculated that
systems corresponding to a large $d$ exhibit a phase transition which is almost
of first order. As mentioned above for finite systems even values $\alpha \leq
0$ cannot be excluded by mathematical reasons.  We note that two-dimensional
Bose-gases are an interesting and growing field of research. As it is well
known, the ideal free Bose-gas in two dimensions ($d=1$) does not show a phase
transition due to thermal fluctuations which destabilize the
condensate~\cite{Mullin97a}.  Switching on a confining potential greatly
influences the properties of the gas, the thermal fluctuations are suppressed
and the gas will show Bose-Einstein condensation.  Recent
experiments~\cite{safonov98} have shown that Bose-Einstein condensation is
possible even though it is called a quasi-condensate. In our notion the
quasi-condensate is just a third order phase transition. Thus, our results are
in complete agreement with recent experiments and earlier theoretical work.  An
interesting question in this respect is whether the order of the transition
changes for $d=2$ in the limit $N\to\infty$. Additional calculation for larger
$N$, which are not printed in Fig.~\ref{analytdimn} indicate that $\alpha$
approaches 1 or might even get smaller. Note that $d=2$ is equivalent to a
hypothetical 4-dimensional ideal Bose gas or Bosons confined in a 2-dimensional
parabolic trap. Our results indicate that the order of the phase transition
sensitively depends on $d$ for values around 2. This might be the reason why
phase transitions in three space dimensions are sometimes classified as second
and sometimes as third order phase transitions.

The parameter $\tau_1$ is a measure of the finite size of the system,
i.e. the scaling behavior of $\tau_1$ as a function of $N$ is a measure
of how fast a system approaches a true n-th order phase transition in
the Ehrenfest sense. The $N$ dependence of $\tau_1$ is displayed in
Fig.~\ref{analytdimn}(c). The scaling behavior can be approximated by
$\tau_1 \sim N^{-\delta}$ with $\delta$ ranging between 1.06 and 1.12
for $d=2-6$.

The $d$ dependence of the classification parameter is visualized 
in Fig.~\ref{analytclassdim} for 50 particles. For this system 
size we found $\alpha \sim d^{-4/3}$ and $\tau_1\sim d^{-4/3}$. 

\begin{figure}[htb]
\centerline{\epsfig{file=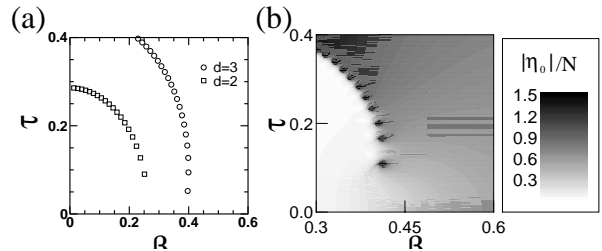,clip=,width=8cm}}
\caption{Comparison between calculated zeros of the canonical partition
function for three dimensional trap geometries with (a) a continuous
single-particle densities of states and (b) discrete energy
levels for $N=40$.}\label{compdens}
\end{figure}

The results presented above for continuous single particle densities 
of states $\Omega(E)=E^{d-1}$ are obtained within semi-analytical 
calculations. In order to compare these results to systems with a
discrete level density we adopt as a reference system the
3-dimensional harmonic oscillator with the
partition function given by 
\begin{equation} \label{garnicht}
Z(\cbeta) = \sum_{n=0} \frac{(n+2)(n+1)}{2} \exp(-\cbeta (n+3/2) ),
\end{equation}
with $\hbar=\omega=k_{\rm B}=1$.

Fig.~\ref{compdens}(a) displays the zeros of the partition
function (\ref{contpart}) for $d=2$ and $d=3$. Fig.~\ref{compdens}(b)
displays a contour plot of the absolute value of the ground state
occupation number $\eta_0(\cbeta) = -\frac{1}{\cbeta}
\partial_{\epsilon_0} Z(\cbeta)/Z(\cbeta)$ with $Z$ given by
(\ref{garnicht}) calculated using an alternative recursion formula
\cite{recurnew}. The zeros of $Z$ are poles of $\eta_0$ and are
indicated by dark spots in this figure.

Analyzing the distribution of zeros consolidates our speculation that the order
of the phase transition sensitively depends on $d$. The distribution of zeros
behaves like the above calculated values for $d=2$ but quantitatively like
$d=3$. Since the degeneracy of the three-dimensional harmonically confined
ideal Bose-gas is a second-order polynomial not only the quadratic term has to
be taken into account.  The linear term becomes dominant for lower
temperatures, so for very low temperatures the best approximation of a
continuous one-particle density of states is $\Omega(E) = E$. The
parameter $\alpha$ supports this statement~\cite{borrmann99}, i.e.~$\alpha$
resides in a region above 1.  Whereas the parameter $\gamma$ behaves like the
$d=3$ case. Finally the parameter $\tau_1$ which is a measure for the
discreteness of the system shows a $\tau_1 \sim N^{-0.96}$ dependence which is
comparable to the one for $d=2$. Thus, for small systems the phase transition is
of third order, it can be speculated if it becomes a second order transition in
the thermodynamic limit.

\begin{figure}[htb]
\centerline{\epsfig{file=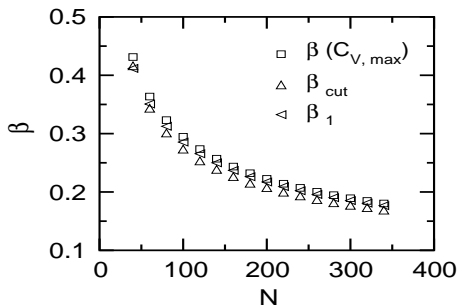,height=4cm,clip=,width=6cm}}
\caption{Comparison between three different approaches to define a critical
temperature for phase transitions in finite systems.}
\label{crittemp}
\end{figure}

Not only qualitatively but also quantitatively our calculations are in very
good agreement with recent theoretical works~\cite{balazs98,Grossmann1995}.
Comparing the {\sl critical} temperature which we defined in Sec.~\ref{Class_sec}
with the usually utilized ones like the temperature of the peak of the specific
heat $\beta({\rm C_{V,max}})$ or the grand-canonically calculated $T_C \sim
N^{1/3}$ confirms our approach. In Fig.~\ref{crittemp} three possible
definitions of the critical temperature are given which all coincide in the
thermodynamic limit. All definitions show a $\beta \sim N^{-\rho}$ dependence
with $\rho$ ranging between $2/5$ and $1/3$. 

\section{conclusion}
Starting with the old ideas of Yang and Lee, and Grossmann {\sl et al.}~we have
developed a classification scheme for phase transitions in finite systems.
Based on the analytic continuation of the inverse temperature $\beta$ into the
complex plane we have shown the advantages of this approach. The distribution
of the so-called Fisher-zeros $\cbeta_k$ draws enlightening pictures even for
small systems whereas the usually referred thermodynamic properties like the
specific heat fail to classify the phase transitions properly. The
classification scheme presented in this paper enables us to name the order of
the transition in a non-ambiguous way. 
The complex parts
$\tau_k$ of the zeros $\cbeta_k$ resemble times for which a whole ensemble of
identical systems under consideration in a heat bath with an initial
Boltzmann-distribution looses its memory. 

We have applied this to ideal non-interacting Bose-gases confined in power-law
traps. We have found that the order of the phase transition sensitively depends
on the single particle density of states generated by the confining potential.
The distribution of zeros exactly reveals the order of the phase
transition in finite systems.

%
%

\end{document}